\documentclass[sigconf]{acmart}
\makeatletter                   
\def\mdseries@tt{m}             
\makeatother                    

\usepackage{amsmath}
\usepackage{graphicx}
\usepackage{booktabs}
\usepackage[inline]{enumitem}
\usepackage{multirow}
\usepackage{adjustbox}
\usepackage{url}

\usepackage[draft=true]{minted}
\usepackage{xcolor}

\usepackage{float}

\usepackage{subcaption}
\usepackage[group-separator={,}]{siunitx}
\usepackage{multirow}

\usepackage{pifont}
\newcommand{\ballnumber}[1]{\tikz[baseline=(myanchor.base)] \node[circle,fill=.,inner sep=1pt] (myanchor) {\color{-.}\bfseries\footnotesize #1};}
\newcommand{\sone}{\ballnumber{1}}
\newcommand{\stwo}{\ballnumber{2}}
\newcommand{\sthree}{\ballnumber{3}}

\usepackage{listings}
\usepackage{courier}

\usepackage{algorithm}
\usepackage{algpseudocode}

\usepackage{tikz}
\usepgflibrary{fpu}

\usepackage{pifont}

\newcommand{\mmm}{M$^3$}

\newcommand{\LL}{\textbf{L$\to$L}}
\newcommand{\LLPr}{\textbf{L$\to$L'}}
\newcommand{\CL}{\textbf{C$\to$L'}}
\newcommand{\LCL}{\textbf{L+C$\to$L'}}

\AtBeginDocument{%
  \providecommand\BibTeX{{%
    \normalfont B\kern-0.5em{\scshape i\kern-0.25em b}\kern-0.8em\TeX}}}

\lstdefinelanguage{CAnDL}
{
morekeywords = { Constraint, with, opcode, collect, include, function\_name, data\_type, ir\_type, domination, strict\_domination, calculated\_from, control\_origin, data\_origin, End },
morecomment  = [s]{\{}{\}}
}

\acmConference[ASE'20]
  {ACM SIGPLAN Conference on Automated Software Engineering}
  {September 21--25, 2020}
  {Virtual}
\acmBooktitle{Proceedings of the 35th International Conference on Automated Software Engineering (ASE '20), September 21--25, 2020, Virtual}
\acmYear{2020}
\acmISBN{} 
\acmDOI{} 
\startPage{1}

\setcopyright{none}

\usepackage{hyperref}
\usepackage{cleveref}
\begin{document}

\title{\mmm{}: Semantic API Migrations}

\author{Bruce Collie}
\orcid{0000-0003-0589-9652}
\affiliation{
  \department{School of Informatics}
  \institution{University of Edinburgh}
  \streetaddress{10 Crichton Street}
  \city{Edinburgh}
  \postcode{EH8 9AB}
  \country{United Kingdom}
}
\email{bruce.collie@ed.ac.uk}

\author{Philip Ginsbach}
\affiliation{
  \institution{GitHub Software UK}
  \city{Oxford}
  \country{United Kingdom}
}
\email{ginsbach@github.com}
\authornote{Work performed while at the University of Edinburgh}

\author{Jackson Woodruff}
\affiliation{
  \department{School of Informatics}
  \institution{University of Edinburgh}
  \streetaddress{10 Crichton Street}
  \city{Edinburgh}
  \postcode{EH8 9AB}
  \country{United Kingdom}
}
\email{J.C.Woodruff@sms.ed.ac.uk}

\author{Ajitha Rajan}
\affiliation{
  \department{School of Informatics}
  \institution{University of Edinburgh}
  \streetaddress{10 Crichton Street}
  \city{Edinburgh}
  \postcode{EH8 9AB}
  \country{United Kingdom}
}
\email{arajan@inf.ed.ac.uk}

\author{Michael F.P.\ O'Boyle}
\orcid{0000-0003-1619-5052}
\affiliation{
  \department{School of Informatics}
  \institution{University of Edinburgh}
  \streetaddress{10 Crichton Street}
  \city{Edinburgh}
  \postcode{EH8 9AB}
  \country{United Kingdom}
}
\email{mob@inf.ed.ac.uk}

\begin{abstract}
  Library migration is a challenging problem, where most existing approaches rely
on prior knowledge. This can be, for example, information derived from
changelogs or statistical models of API usage.  

This paper addresses a different API migration scenario where there is no prior
knowledge of the target library. We have no historical changelogs and no access
to its internal representation. To tackle this problem, this paper proposes a
novel approach (\mmm{}), where probabilistic program synthesis is used to
\emph{semantically} model the behavior of library functions.  Then, we use an
SMT-based code search engine to discover similar code in user applications.
These discovered instances provide potential locations for API migrations.

We evaluate our approach against 7 well-known libraries from varied application
domains, learning correct implementations for 94 functions. Our approach is
integrated with standard compiler tooling, and we use this integration to
evaluate migration opportunities in 9 existing C/C++ applications with over
1MLoC. We discover over 7,000 instances of these functions, of which more than
2,000 represent migration opportunities.

\end{abstract}


\maketitle

\section{Introduction} \label{sec:intro}
\subsection{API Migration}

Libraries are a fundamental feature of software development. They allow the
sharing of common code, separation of concerns and a reduction in overall
development time. However, libraries are not static. They continually evolve to
provide increased functionality, security and performance. Unfortunately,
upgrading software to match library evolution is a significant engineering
challenge for large code bases.
 
Given the wide-scale nature of the problem, there is much prior work in the area
under various headings (e.g.\ library upgrade, API evolution or library
migration). Work in these areas aims to answer the same question: when (and how)
can a program using API $X$ be transformed to one that uses API $Y$ while
preserving its behavior? This is a difficult problem even when $X$ and $Y$ have
similar interfaces. It becomes more challenging if their behaviors do not match,
and requires surrounding code to be factored in.

There are several approaches to this migration problem: if examples exist of
previous successful migrations, then these examples can be used to derive
mapping rules \cite{Teyton2013a}. This approach requires that a full history of
the application's source code is available, annotated with the libraries in use
at each commit. Neural models have been used successfully to predict properties
of programs based on learned vector-space embeddings \cite{nguyen2017exploring}.
However, these approaches require large training sets and are imprecise with
respect to program semantics. A more precise (but less automatic) approach is to
use expert knowledge to encode known migration patterns
\cite{Wasserman2013,Savchenko2019}.

All these prior approaches require some knowledge of the API. In this paper we
tackle the challenging task of API migration \emph{without} any prior knowledge
of the source or target libraries. Here, we do not have access to the library's
source code, nor to a corpus of example usages of the library. While this
scenario may seem draconian, it is often the case in practice
\cite{kuznetsov2010testing}. Libraries may be closed-source \cite{tang2020libdx}
or distributed in binaries for convenience \cite{Miranda2018}, and could even be
implemented as hardware \cite{coelho2013api}. In this paper we propose a novel
approach which automatically learns pattern-based semantic migrations, but
without up-front expert knowledge. 

\begin{figure*}[t]
  \includegraphics[width=\textwidth]{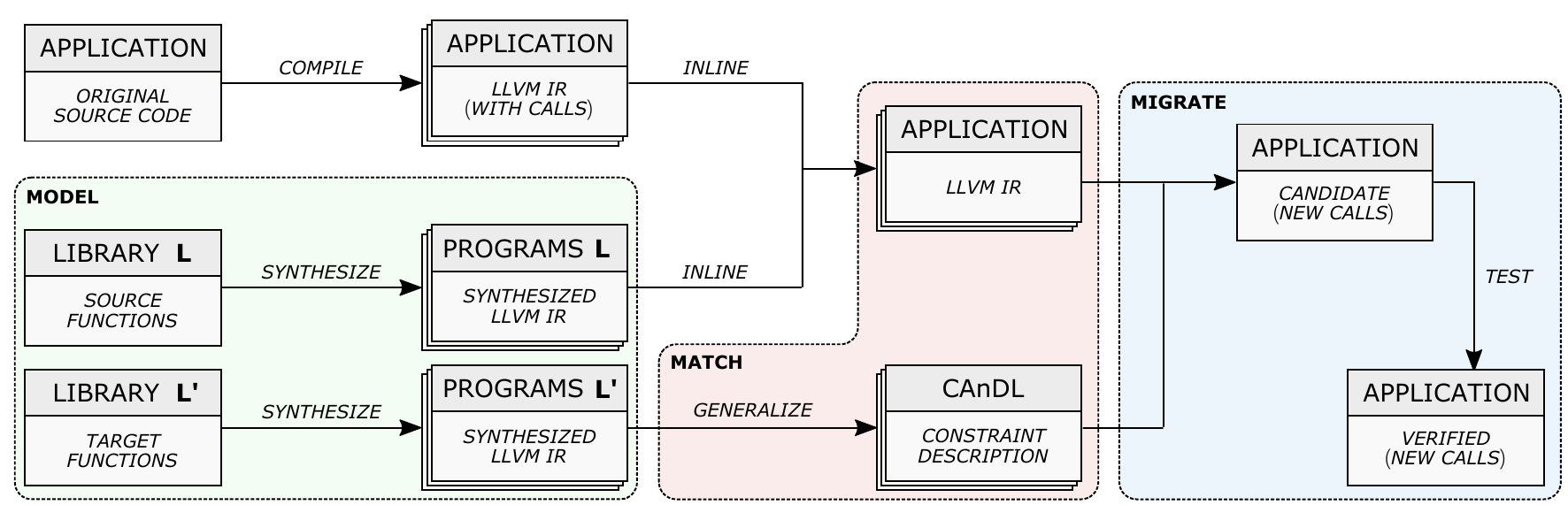}

  \caption{
    A summary of the \mmm{} workflow. Models for library functions are
    synthesized. Source functions are inlined while synthesized target functions are
    generalized into constraint descriptions, which are then used to search
    compiled user code for potential migrations.
  }

  \label{fig:system}
\end{figure*}

\subsection{\mmm{}: Model, Match and Migrate}

The key to our approach is to derive a model that is actual executable code. We
call such an approach \emph{semantics-based} migration. Given a specification
for a library function (type signature, function name, library binary containing
its implementation), \mmm{} attempts to automatically \emph{Model} its behavior
using program synthesis and checks correctness with respect to automatically
generated input-output examples. It inlines the learned program models, then
uses compiler-based constraint analysis to \emph{Match} regions of application
code with compatible libraries. Finally, we \emph{Migrate} these regions by
replacing application code with library calls.

A useful feature of this approach is that as well as library migration, it
allows the refactoring of library-free user code to use libraries. This is
because the synthesized models are themselves code, and are inlined and analyzed
together with application code. Complex refactorings that integrate contextual
code around an API call are enabled by this approach.

Our approach, while having the benefit of not requiring library vendors to
release their source code, relies on the ability to synthesize programs in a
reasonable time. We build on methods from sketch-based synthesis
\cite{Solar-Lezama2009} to discover program structure before performing a
directed enumerative search. We incorporate new probabilistic models to more
effectively navigate the large search space. We evaluate our approach across 7
libraries, synthesizing 94 functions, and match them to over 7,000 old library
calls across 10 applications with up to 1MLoC. We were able to successfully
migrate more than 2,000 of these calls to another library.

\paragraph{Summary of Contributions}

We provide a novel and efficient program synthesizer for real-world library
functions. Additionally, we detail a method for matching similar code in
applications using solver-aided techniques. Using this procedure, we are able to
discover opportunities for usages of a source library \textbf{L} or application
code \textbf{C} to be migrated to a target library \textbf{L'}. Furthermore, we
can migrate source libraries with surrounding contextual code (\textbf{C + L})
to target libraries. This is achieved \emph{without any knowledge} about the
implementation of either library.

\section{Overview} \label{sec:overview}

In this section, we first present a high-level summary of the \mmm{} workflow,
then show an example of the type of migrations it enables in practice.

\subsection{\mmm{} Workflow}

\Cref{fig:system} shows the flow of data through \mmm{}. It takes as input an
application, along with specifications for source and target library APIs
(currently used \textbf{L} and potential targets \textbf{L'}). The end result is
a modified application that references the target libraries. We highlight the
three distinct phases: \textbf{Model}, \textbf{Match} and \textbf{Migrate}.

\subsubsection{Model}

We assume that the source code for libraries is not available, as is often the
case in practice \cite{kuznetsov2010testing}. The first phase of \mmm{} is
Model: the synthesis of programs equivalent to functions in both the source and
target libraries. The programs we synthesize are in the form of LLVM
\cite{Lattner2002} intermediate representation; this representation allows us to
directly integrate synthesized programs in existing compiler toolchains, and to
benefit from robust program manipulation libraries. The synthesis process is
specified using randomly-generated input-output examples (see
\Cref{sec:randomgen}). 

\subsubsection{Match}

The second phase, Match, uses the synthesized implementations of source and
target library functions in two ways. First, we inline the synthesized code of
the \emph{source} library functions into the user application at each call site.
Secondly, we generalize the synthesized code of the \emph{target} library
functions to a constraint-based description that allows for matching code to be
efficiently searched for.

Performing inlining means that the behavior of the library function and the
\emph{context} in which it appears are unified; migrations that require
splitting, merging or moving functionality can be discovered and performed.

\subsubsection{Migrate}

Once matches are found, we verify whether or not potential migrations are
correct. First, we perform basic integration testing using random examples on
the new code. This helps to eliminate false positive matches. At this stage, the
migration can be performed automatically, although in practice the user would be
asked to confirm the migration (as is usual with API migration tools). We
perform integration testing to check correctness of Migrate.

\subsection{Example}

\begin{figure}
  \centering
  \includegraphics[width=0.9\columnwidth]{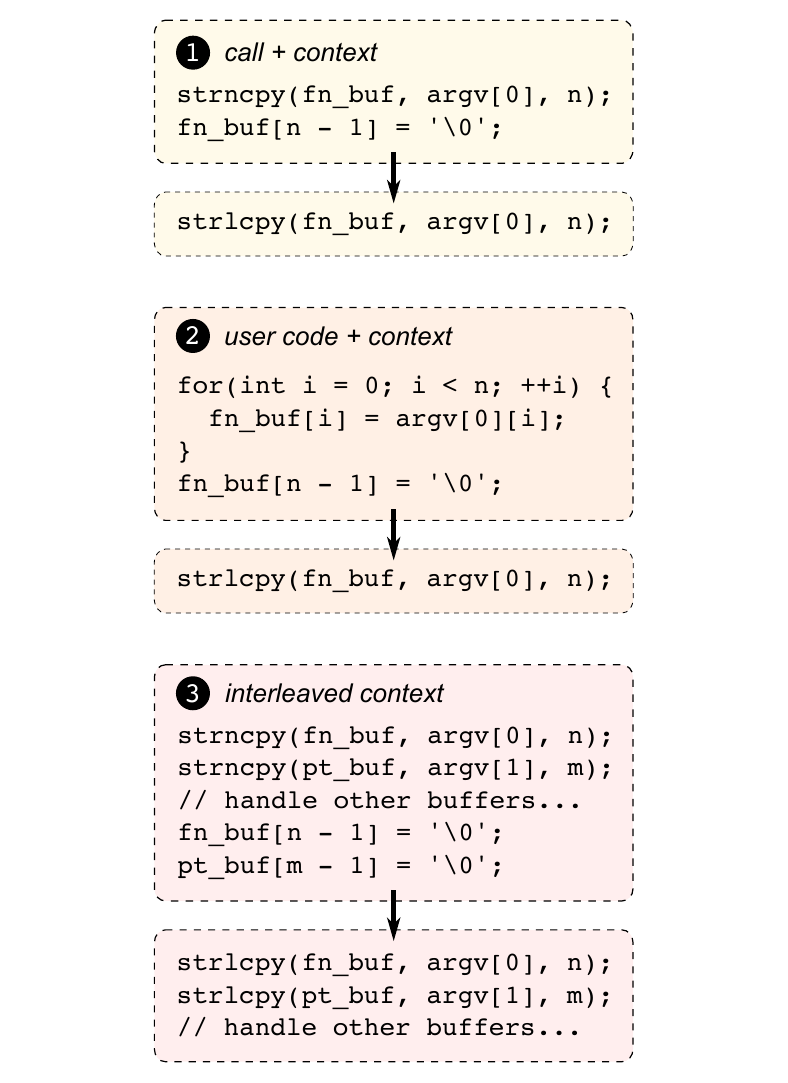}

  \caption{
    Example of three contexts in which \mmm{} is able to perform contextual API
    migrations using only the behavior of the source and target functions.
  }

  \label{fig:example}
  \vspace{-10pt}
\end{figure}

To demonstrate the types of migration that \mmm{} offers, we use a running
example taken from the Common Weakness Enumeration (CWE) database \cite{CWE}. If
the standard \mintinline{C}{strncpy} function is used to copy a C string,
null-termination is not guaranteed. This can lead to buffer over-reads, and so
alternative functions often exist to perform a terminated copy (for example,
\mintinline{C}{strlcpy} on BSD, \mintinline{C}{StringCchCopy} on Windows or
application-specific implementations). CWE-126 identifies a common pattern of
manually adding string terminators that can be replaced by these functions;
doing so is a useful API migration task.

\Cref{fig:example} shows the three patterns identified in CWE-126 that can be
refactored for safety. The first case \sone{} is the simplest: a call to
\mintinline{C}{strncpy} is immediately followed by an explicit termination. This
migration could be performed using tools such as Refaster \cite{Wasserman2013},
but would require an expert to encode it manually.

The second case \stwo{} highlights the utility of \mmm{}: after performing
inlining, the code that explicitly calls \mintinline{C}{strncpy} is no different
to code that performs an explicit loop. Both of these patterns exist in real
code, and can be migrated equivalently using \mmm{}. Because many different
syntaxes might represent the same semantics, writing source-code based tools
that discover loops in this way is a hard problem \cite{Ginsbach2018a}; \mmm{}'s
compiler integration and IR-level search allows it to handle loops and other
control flow statements seamlessly.

Finally, the third case \sthree{} shows a complex migration where calls to
\mintinline{C}{strncpy} are interleaved with their respective terminations. By
operating at the IR level, \mmm{} is able to identify that no dependencies exist
between the calls, and so the migration is possible. In general, source
code-based tools, even with expert knowledge, are less able to make this
determination.

Unifying these different forms of migration without requiring up-front expert
knowledge or library source code is the key advantage of \mmm{}.

\section{Model} \label{sec:model}
The Model phase of \mmm{} is a program synthesizer; it aims to generate
functions that behave equivalently to target library functions. Model uses
component-based sketching \cite{Jha2010} together with novel learned
probabilistic models to efficiently search for the most likely structure for
correct solutions. Then, a gradual refinement process is used to instantiate
working programs from these structures. Candidates are tested against the target
function using randomly-generated inputs; the adequacy of this testing strategy
is validated using branch coverage.

\subsection{Correctness}

Providing a formal proof of total correctness for this type of synthesis problem
is extremely complex~\cite{deville1994logic}.  In this paper, we define
correctness using the standard formulation of \emph{observational equivalence}:
a candidate is correct if it behaves identically to the target over a particular
set of input examples. Most work in synthesis using input-output examples shares
this formulation \cite{So2017,David2017}.

This definition relies on a good enough set of input examples being available.
We cannot rely on the user knowing enough of the target's semantics to produce a
set of minimal, interesting examples \cite{Nye2019,Feser2015} (and in fact wish
to abstract this process away from the user). We therefore resort to random
generation of input examples.

\subsubsection{Generating Test Inputs} \label{sec:randomgen}

The Model phase supports the primitive C types \mintinline{C}{char},
\mintinline{C}{int} and \mintinline{C}{float}, and pointers to these types.
Values of integer and floating-point types are generated by sampling values
uniformly in the range $ [-64, 64] $, and for characters from their entire
range. For pointer data, blocks of 4,096 elements are allocated (to allow for
large computed indices based on input data). Each element of these blocks is
sampled according to the appropriate scalar sampling method. 

Existing work on fuzzing and automated testing \cite{Zalewski2020} generally
observes that interesting behavior most often occurs at small input values; our
input range was selected to provide a varied distribution of values while also
maintaining a reasonable probability of generating small (and therefore
interesting) inputs.

Our input generation methodology can be easily generalized to more types;
further primitive types (e.g.\ differently sized integers or booleans) follow
the same methodology, while aggregate types (e.g.\ a C \mintinline{C}{struct})
can be generated compositionally over their individual elements. Generating data
structures with internal invariants or unusual distributions is an open problem
\cite{Shin2019}.

\subsubsection{Testing Coverage} 

It is important that the randomly generated inputs properly exercise the
possible behaviors of both the target and candidate. While it is not possible to
measure coverage for a black-box target in the absence of source code, we
measure \emph{branch coverage} over each candidate during synthesis. New inputs
are generated until full coverage is achieved.

Our results in \Cref{sec:rq2} show that random testing and coverage measurement
is an effective means to validate the behavior of synthesized programs. 

\subsection{Specification}

Two inputs fully specify a synthesis problem: the type signature and name of the
target function, and a library containing an implementation with that name.
There are no requirements on the internal details of this implementation.

No further information about the target function is required. For example, base
cases or semantic annotations (such as those used by $\lambda^2$
\cite{Feser2015} or in the type-directed synthesis procedure demonstrated by
\citet{Collie2019}) are not required by our implementation, and we do not
require manually created inputs to test candidate programs as other synthesizers
such as \textsc{Simpl} \cite{So2017} or \textsc{SketchAdapt} \cite{Nye2019} do.

\subsection{Fragment-Based Sketching}

Program synthesis commonly divides the search for a solution into two phases.
The first, sketching, aims to establish the \emph{structure} of a solution. In
its initial formulation, sketches were provided by the user based on their
insight into the problem \cite{Solar-Lezama2009}. By doing so, search for
programs with complex control flow could be reduced to more tractable problems.
More recent approaches aim to synthesize the sketch as well
\cite{Nye2019,Wang2017,Ellis2018}. Our approach falls into this group as it does
not require the user to provide any sketch information. Instead, it uses a novel
probabilistic approach.

We aim to build sketches compositionally from smaller \emph{fragments}, which
represent independent elements of program structure. For example, a program that
performs a linear search may comprise a loop fragment composed with a
conditional test fragment. Some fragments are parameterized; in these cases
different variants of the fragment are instantiated depending on the available
information for a given problem. The full set of fragments used by \mmm{} to
perform synthesis are listed below, along with C-like pseudocode describing
their semantics.

\subsubsection{Fragment Corpus}

The library of fragments used by Model is given below. The \mintinline{C}{use}
function represents code generated that may use a particular value,
\mintinline{C}{?} is possible composition, and \mintinline{C}{_P} is a
placeholder value of appropriate type.

\pagebreak

\begin{description}
  \item[Linear] A basic block into which instructions should later be
    synthesized.

  \item[Fixed Loop] Template for a loop with known upper bound, parameterized on
    an optional pointer \mintinline{C}{ptr} and an integer \mintinline{C}{x}:
\begin{minted}{C}
for(int i = 0;i < x;++i) { ? }
for(int i = 0;i < x;++i) { use(ptr[i]);? }
\end{minted}

  \item[Delimiter Loop] Template parameterized on a pointer \mintinline{C}{ptr}:
\begin{minted}{C}
while(*ptr++ != _P) { use(*ptr); ? }
\end{minted}

  \item[Loop] A catch-all for iterations not covered by the two more specialized
    fragments:
\begin{minted}{C}
while(_P) { ? }
\end{minted}

  \item[If, If-Else] Conditional control flow:
\begin{minted}{C}
if(_P) { ? }
if(_P) { ? } else { ? }
\end{minted}

  \item[Seq] Execute two fragments, one after the other:
\begin{minted}{C}
? ; ?
\end{minted}

  \item[Affine, Index] Synthesize affine and general index expressions
    respectively, parameterized on \mintinline{C}{ptr}. For example:
\begin{minted}{C}
int a_v = ptr[_P * _P + _P]; // e.g.
int v = ptr[_P - _P];        // e.g.
\end{minted}

\end{description}

The available set of fragments for a synthesis problem depends on which ones can
be properly instantiated; we write $ \mathbf{F} $ for this set. In this work we
use only the fragments described above, but it is possible for users to extend
the corpus of fragments (for example, to specialize for a particular problem
domain with partially-known structure).

\subsubsection{Composition} 

We define an intuitive composition operation between any two fragments, with $
\circ $ the left-associative composition operator.

\subsection{Probabilistic Models}

The set of possible fragment compositions for some problems is very large. Model
uses two cooperating probabilistic models to reduce the size of the search
space. The first predicts which fragments from the available set are most likely
to appear anywhere in a correct solution, and the second uses a Markov model to
identify compositions of fragments most likely to yield a correct program.

\subsubsection{Fragment Likelihood} 

We use a random forest classification model to predict, for each fragment $ f
\in \mathbf{F} $, whether it appears in a correct solution program. The
classifier $ C $ takes as input a fragment $ f $ and type signature $ \tau $,
and outputs a prediction of whether this fragment will appear in a correct
solution for a function with that type signature. Applying the classifier $ C $
to every fragment produces a predicted set of fragments $ \mathbf{F_0} \subseteq
\mathbf{F} $:
\[
  \mathbf{F_0} \triangleq \{ f \in \mathbf{F} \;|\; C(f, \tau) \}
\]
We achieved a mean Jaccard score of 0.82 between $ \mathbf{F} $ and $
\mathbf{F_0} $ using this predictor; this means that the predictor does not
significantly over- or under-approximate.

\subsubsection{Composition Sampling}

There are a large number of potential compositions over $ \mathbf{F_0} $ that
produce a sketch. It is therefore important to predict which compositions are
the most likely to produce a correct solution.

We equate the linear sequence of fragments $ f_1, f_2, \dots, f_n $ with the
composition $ f_1 \circ f_2 \circ \dots \circ f_n $. This allows us to sample
compositions using a simple Markov model. To do so, we add start and end symbols
to the fragment vocabulary, and sample fragments according to the probability $
\mathbb{P}(f_{n} | f_{n-1}) $ until the end symbol is sampled.

The conditional probability $ \mathbb{P}(f_n | f_{n-1}) $ is trained using
observed fragment composition bigrams. For example, if a sketch from composition
$ f \circ g \circ h $ produces a correct solution, then the bigrams $ f \circ g
$ and $ g \circ h $ are both observed. Based on a matrix $ w $ of observation
counts (where $ w(f, f') $ is the observed count of $ f \circ f' $), we define:
\[
  w'(f_i, f_j) \triangleq
  \begin{cases}
    w(f_i, f_j) & \text{if } f_j \in \mathbf{F_0} \\
    0 & \text{otherwise}
  \end{cases}
\]
\[
  s(f) \triangleq \sum_{f\in \mathbf{F_0}} w(f, f')
\]
\[
  s'(f) \triangleq \sum_{f\in \mathbf{F_0}} w'(f, f')
\]
Then, the Markov probabilities can be given as:
\[
  \mathbb{P}(f_n | f_{n-1}) \triangleq b \frac{w'(f_n,
  f_{n-1})}{s'(f_{n-1})} + (1 - b) \frac{w(f_n,f_{n-1})}{s(f_{n-1})}
\]
where $ b \in [0, 1] $

\subsubsection{Training} 

To train these models, a 25\% subset of our evaluation library functions was
selected randomly. Ground truth sketches were constructed by hand for each
target function in this subset and used to train both models. A 25\% training set
split was identified through manual parameter search; there is enough redundancy
among the functions to enable the use of a small training proportion. 

We do not believe the construction of such a training dataset to be particularly
onerous when compared to statistical migration techniques, which often entail
cleaning and preprocessing millions of lines of code. Additionally, it is
possible to \emph{bootstrap} our training set starting from the synthesized
solutions to simple problems. Doing so provides no benefit to the model
performance (only to the collection of training data), and so in this paper we
do not examine the process.

\subsection{Instruction Search}

The final step in Model's synthesis process is to perform an enumerative search
for candidate programs based on predicted sketches. Each fragment specifies a
set of typed placeholder values; these identify where computation can be
performed within that fragment. For example, in the LLVM code below, the values
\mintinline{llvm}{
values of type \mintinline{llvm}{i32}. Placeholders may also be untyped
(\mintinline{llvm}{

\begin{minted}{llvm}
%0 = call i32 @ph_i32()
%1 = call void @ph()
%2 = call i32 @ph_i32()
%3 = call i1 @ph_i1()
\end{minted}

To search for candidate programs based on a sketch with placeholders, Model
assigns concrete values to each placeholder in turn; different choices of values
produce different programs. As values are selected, the potential choices for
other values may be restricted. The result of this is a lightweight
constraint-solving problem (for example, if the value \mintinline{llvm}{add i32
\mintinline{llvm}{
  details of how this constraint problem can be implemented by the compiler are
  given in prior work \cite{Collie2020}. To optimize the traversal of a
  potentially large search space, Model uses the following heuristics: 

\begin{itemize}
  \item Placeholders of known type are assigned first.
  \item When selecting operands for unary or binary operators, operands located closer
    to the operator are prioritized.
  \item More common operators are prioritized (e.g.\ addition is attempted
    before division).
  \item A threshold for the total number of instructions is set and iteratively
    relaxed (e.g.\ initially programs of 3 concretized instructions are
    considered, then 4 if no successful candidate is found, etc.).
\end{itemize}

By assigning values in this way, a concrete program is gradually refined from a
sketch. Fragments are not required to enforce concrete types on their
constituent values, but can enforce constraints when they do (e.g.\ a
conditional fragment requires a boolean value).

Once we have a complete program, we can compile and execute it using randomly
generated input values.

\section{Match} \label{sec:match}
Once Model has synthesized a program with behavior equivalent to a target
library function, the next step is Match: we aim to discover regions of code
that are equivalent to the synthesized implementation.

\subsection{Searching for Code Using CAnDL}

Efficiently searching for sections within an application that satisfy particular
criteria is a hard problem to express using traditional programming languages.
The CAnDL language \cite{Ginsbach2018a} allows for declarative specification of
search patterns, which are compiled to constraint-satisfaction problems that can
be efficiently resolved using backtracking search (in a manner similar to SMT
solvers \cite{Barrett2018}).

CAnDL patterns specify dataflow relationships between values in LLVM IR
programs, as well as properties of individual values. For example, the property
``$x$ is an add instruction, and $y$ is a multiply with $x$ as one of its
operands'' is a simple CAnDL pattern. These patterns amount to a set of
\emph{constraints} on the program that must be satisfied for the pattern to
match it; searching for matching code is therefore a constraint satisfaction
problem.

We use the standard CAnDL toolchain to efficiently solve such constraint
problems over LLVM. Full details of the search algorithms can be found in
\cite{Ginsbach2018a}; in this paper we take as given an efficient solution
procedure for CAnDL-compatible constraint problems over LLVM IR programs.

\subsection{Translating LLVM to Constraints}

In \cite{Ginsbach2018a} the authors write CAnDL constraints by hand to match
specific computational idioms (for example, polyhedral control flow or stencil
codes) of interest to a domain-specific optimizer. By comparison, we aim to
generate constraints automatically from synthesized code. Automatic generation
of constraints is not a use case envisioned by the original authors, and so we
contribute a novel algorithm for emitting constraint descriptions from example
LLVM programs.

\Cref{fig:llvmfrag} shows a small fragment of LLVM IR; the code is in SSA form
and can be described by a directed acyclic graph. Below, \Cref{fig:candlfrag}
gives a set of CAnDL constraints that describe this fragment. Each instruction
(as well as constants and function parameters) occurs as a variable name in the
constraints; the constraint program serves as a description of the data flow.
The data flow graph is serialized by classifying individual variables (lines
2--10), and then the interactions between them (lines 11--21). This description
is passed to the CAnDL solver to efficiently find satisfying code.

\definecolor{mblue}{rgb}{0.27,0.33,0.53}
\definecolor{mgreen}{rgb}{0.27,0.53,0.33}
\definecolor{mred}{rgb}{0.63,0.23,0.27}
\newcommand{\opcode}{\textcolor{mblue}{\textbf{opcode}}}
\newcommand{\type}[1]{\textcolor{mgreen}{\textbf{#1}}}
\newcommand{\irtype}{\textcolor{mred}{\textbf{ir\_type}}}

\begin{figure}[t]
  \begin{subfigure}[b]{\columnwidth}
    \begin{minted}{llvm}
%iter = phi i64 [%new_iter,%loop], [0,%entry]
%addr = getelementptr i64,
          i64* %array, i64 %iter
%elem = load i64, i64* %addr
%niter = add i64 %iter, 1
    \end{minted}
    \subcaption{
      Fragment of LLVM code extracted from a function that computes the sum of
      an array of integers.
    }
    \label{fig:llvmfrag}
  \end{subfigure}

  \bigskip

  \begin{subfigure}[b]{\columnwidth}
    \begin{minted}[xleftmargin=2em,linenos,escapeinside=||]{text}
|\textbf{Constraint}| Generated
( |\opcode{}|{iter}     = |\type{phi}|
& |\opcode{}|{addr}     = |\type{gep}|
& |\opcode{}|{elem}     = |\type{load}|
& |\opcode{}|{niter}    = |\type{add}|
& |\irtype{}|{0}       = |\type{literal}|
& |\irtype{}|{1}       = |\type{literal}|
& {niter}          = {iter}.arg[0]
& {0}              = {iter}.arg[1]
& {array}          = {addr}.arg[0]
& {iter}           = {addr}.arg[1]
& {addr}           = {elem}.arg[0]
& {iter}           = {new_iter}.arg[0]
& {1}              = {new_iter}.arg[1])
|\textbf{End}|
    \end{minted}
    \subcaption{
      CAnDL constraints generated from the LLVM code above.  These
      constraints capture the structure of the code and can be
      efficiently searched for in large LLVM code bases.
    }
    \label{fig:candlfrag}
  \end{subfigure}

  \caption{
    LLVM code sample and its corresponding CAnDL constraints, as
    generated by Match.
  }
  \label{LLVMconstraintsexample}
\end{figure}

Our constraint descriptions are built from a dataflow graph representation of
LLVM IR, where vertices are instructions and edges capture the argument
relation. \Cref{firstalgemit} shows how we generate a description of this graph
structure.

Looping over the graph vertices (lines 4--17), the instruction opcode
constraints are emitted, as well as the constraints that deal
specifically with constant and function argument values. In a second
loop (lines 18--20), the data flow graph is serialized by iterating over
the graph edges and emitting positional argument constraints. The
remaining lines of the algorithm generate the logical conjunctions
holding the individual constraints together (lines 5--9) and produce the
boilerplate CAnDL code (lines 2 and 21).

\subsection{Post-Processing Constraints}

This approach results in a constraint program that searches for exact sub-graph
matches in user code, but is often too specific. We therefore apply a careful
weakening of the constraints to produce a more general matching.

Firstly, constraints that specify values to be function arguments are
counterproductive; these constraints will not hold after inlining, so they are
removed in post-processing. Secondly, some operators are commutative and
therefore the positional argument constraints on them are too strict. They are
replaced with a logical disjunction between the corresponding permutations.
Finally, we remove instructions that correspond only to compiler-specific code
generation idioms.

\begin{algorithm}[t]
  \caption{Emit Constraint Description}
  \label{firstalgemit}
  \begin{algorithmic}[1]
    \Function{EmitConstraints}{$V$,$E$}
      \State\textsc{emit}("Constraint Generated (")
      \State $first \leftarrow true$
      \For{$ v $ in $ V $}
        \If {$first$}
            \State $first \leftarrow false$
        \Else
          \State\textsc{emit}("\&")
        \EndIf
        \If {$op(v) = \text{parameter}$}
            \State\textsc{emit}("ir\_type", $name(v)$, " = argument")
        \ElsIf {$op(v) = \text{const}$}
            \State\textsc{emit}("ir\_type", $name(v)$, " = literal")
        \Else 
            \State\textsc{emit}("opcode", $name(v)$, " = ", $op(v)$)
        \EndIf
      \EndFor
      \For{$n,a,b $ in $ E $}
         \State\textsc{emit}($name(a)$, " = ", $name(b)$, ".args[", $n$, "]")
      \EndFor
      \State\textsc{emit}(") End")
    \EndFunction
  \end{algorithmic}
\end{algorithm}

\begin{table}[t]
    \centering
    \caption{Corpora used to evaluate \mmm{}.}
    \label{tab:corpora}

    \subcaption{Application source code for which migrations were tested.}
    \label{tab:sources}
    \begin{tabular}{l|lr}
      \toprule
      \textbf{Software} & \textbf{Description} & \textbf{LoC} \\
      \midrule
      \texttt{ffmpeg} & Media processing & 1,061,655 \\
      \texttt{texinfo} & Typesetting & 76,755 \\
      \texttt{xrdp} & Remote access protocol & 75,921 \\
      \texttt{coreutils} & Utilities & 66,355 \\
      \texttt{gems} & Graphics helpers & 46,619 \\
      \texttt{darknet} & Deep learning & 21,299 \\
      \texttt{caffepresso} & Deep learning & 14,602 \\
      \texttt{nanvix} & Operating system & 11,226 \\
      \texttt{etr} & Game & 2,399 \\
      \texttt{androidfs} & Filesystem & 1,840 \\
      \bottomrule
    \end{tabular}
    \vspace{2em}
    \subcaption{Library APIs for which synthesized implementations were learned
    and used to drive migration.}
    \label{tab:libraries}
    \begin{tabular}{l|l}
    \toprule
    \textbf{Library} & \textbf{Description} \\
    \midrule
    \texttt{string.h} & C standard library string handling \\
    \texttt{StrSafe.h} & Safety-focused C string handling \\
    \texttt{glm} & Graphics functions \\
    \texttt{mathfu} & Mathematical functions \\
    \texttt{BLAS} & Linear algebra \\
    \texttt{Ti DSP} & DSP Kernels \\
    \texttt{ARM DSP} & DSP Kernels \\
    \bottomrule
    \end{tabular}
\end{table}

\section{Migrate} \label{sec:migrate}
Model and Match make up the bulk of the work done by \mmm{}. The final step is
to leverage the synthesized programs and generated constraints to generate
appropriate API migrations. We produce source-level substitutions that can be
applied manually to the source code, as well as automatically-tested IR
transformations.

\subsection{IR-level Replacements}

Migrate is able to automatically apply a potential migration within an
application being compiled. To do this, the IR values that matched against a
library function's parameters and return value are identified. A call to the
function is inserted with the appropriate arguments given, and uses of
the matched return value are replaced with the new call's return value.

\pagebreak

By doing this, we obtain a modified version of the application's code. Regions
that match the generated constraints for library functions are replaced with
calls to those library functions. Migrate extends the functionality used in the
original CAnDL paper \cite{Ginsbach2018a} by not requiring the migration process
to be implemented manually for every relevant library function; having the
synthesized code available to map values allows us to do this.

\subsection{Validation} \label{sec:validate}

The primary usage of automated IR replacement is to validate migrations (i.e.\
to check whether or not performing the migration will result in a correct
program). While formally proving this is unlikely to be possible for any API
migration tool, Migrate performs two validation steps that provide some
assurance that its suggestions are correct. First, we ensure that no
dependencies to intermediate values in the pre-replacement code exist later in
the function. Then, we test the code post-replacement with random IO examples
using the same methodology as Model uses; our results in \Cref{sec:rq3results}
show that this validation is effective.

Beyond these checks, the user is likely to still perform their own validation
(e.g.\ running unit or integration tests). Other API migration tools share this
characteristic; no changes suggested by refactoring tools to any codebase are
likely to go untested.

\subsection{Source-level Suggestions}

Our methodology for this paper operates at the IR level, within the compiler;
migrations are applied mechanically by performing substitutions of SSA values.
Doing so allows us to automatically test applied migrations, but changes made at
the IR level can be difficult for a user to understand.

We implemented a prototype tool that used LLVM's debugging libraries to generate
source-level suggestions instead. Source-level suggestions are harder to apply
mechanically, but allow for easier user insight into what changes have been made
by the migration. Evaluating this tool is outside the scope of the paper (as its
usage was not necessary for any of our research questions), but we hope to
implement it more fully and perform a user study as future work.

\section{Experimental Design} \label{sec:setup}
To evaluate the success of \mmm{}, we identify four research questions:
\begin{description}

  \item[(RQ1) Feasibility and effectiveness of the Model phase:] Can program
    synthesis be used effectively to learn the behavior of black-box library
    functions?

  \item[(RQ2) Correctness of synthesized programs: ] Do the synthesized programs
    behave the same as the target program over a particular set of inputs?  The
    inputs used for this correctness check are randomly generated. To assess the
    adequacy of the random inputs in checking behaviors of the synthesized and
    target programs, we measure branch coverage achieved by the random inputs
    over them.   

  \item[(RQ3) Accuracy of Match phase:] Given synthesized implementations for
    library functions, can compatible instances in application code be
    accurately discovered? In this research question, we focus on ability and
    accuracy of the Match phase to discover inlined implementations of the
    \emph{same} synthesized library functions in application code.

  \item[(RQ4) Accuracy of Migrate phase:] Given instances of user code that
    match the constraints generated from a library function, can API migrations
    be correctly implemented? This research question investigates ability and
    accuracy of the Migrate phase in matching and migrating implementations in
    application code to \emph{different} library functions. 

\end{description}

\subsection{Evaluation Corpora}

\subsubsection{Applications}

We selected 9 widely-used applications to evaluate our approach against; they
are listed in \Cref{tab:sources}. Each application is written in C or C++, and
they cover a wide range of problem domains.

We selected these applications by manually searching GitHub and similar online
repositories\footnote{Using \url{https://searchcode.com/}} for code that matched
the following criteria: most importantly, we required a build system that
permitted easy interposition of our compiler toolchain. For our purposes, this
ruled out applications not written in C or C++, although with some additional
engineering work any language with an LLVM frontend could be integrated. 

When selecting applications, large, popular and real-world code was prioritized.
We selected projects in active development, or those for which significant
distribution and usage could be identified. We aimed for a diverse range of
application domains with minimal duplication. No pre-selection of applications
based on knowledge of their source code was performed; the authors were not
familiar with these applications in advance.

\subsubsection{Libraries}

We selected 7 libraries to target for migration, from two broad problem domains:
string processing and mathematical operations. Similar domains are commonly
targeted by other migration tools (with different tooling and language
contexts). We required libraries that could be called easily from C/C++ for
compatibility with the synthesizer. 

For string processing, our starting point was the standard C \texttt{string.h}
header along with BSD extensions. We additionally selected the Microsoft
\texttt{StrSafe.h} library that extends the standard functions with safer
alternatives that avoid common security issues. We then selected five
mathematical libraries with slightly different areas of application and platform
support in order to evaluate the ability of \mmm{} to discover cross-vendor or
cross-platform migrations. Other work \cite{Collie2019} identifies the
usefulness of this type of migration. Full details of the selected libraries are
given in \Cref{tab:libraries}.

\section{Results} \label{sec:results}
\subsection{RQ1: Feasibility and Effectiveness of Model} \label{sec:rq1}

\begin{figure}[t]
  \centering
  \includegraphics[width=\columnwidth]{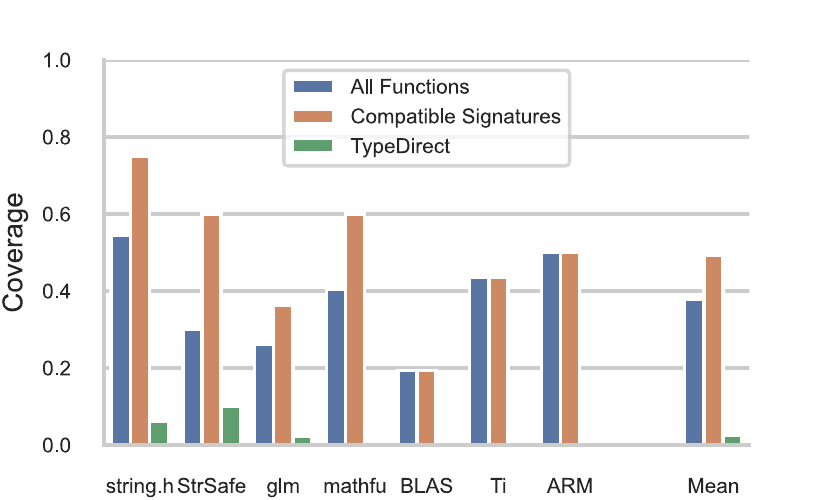}

  \caption{
    Proportion of each library's API that we were able to successfully
    synthesize, across all functions in the library as well as those with
    compatible type signatures. Results are also shown for \textsc{TypeDirect}
    \cite{Collie2019}.
  }

  \label{fig:coverage}
\end{figure}

\begin{figure}[t]
  \centering
    \includegraphics[width=\columnwidth]{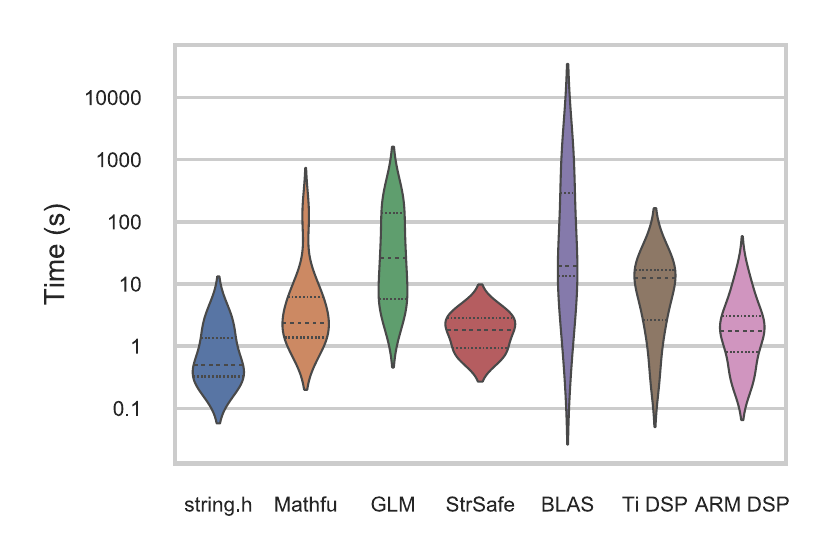}

    \caption{
      Distribution of synthesis times for each library API.
    }

    \label{fig:difficulty}
\end{figure}

\subsubsection{Library Coverage}

\Cref{fig:coverage} shows the proportion of each library's API we were able to
synthesize correctly across all functions in the library (shown as blue bars).
As expected, we could not synthesize every function from each library. The
primary reason for a synthesis failure was a function's type signature not being
compatible with Model, for example those using pointers to pointers or complex
structure types. Beyond these failures, there were a number of cases where
internal data structure usage meant that Model's control flow fragments were not
able to express the necessary structure (e.g.\ the control flow required to
operate on the packed matrices in \texttt{strsm} from BLAS). 

Model successfully synthesizes implementations for an average of 37\% of the
functions in each library evaluated (blue bars in \Cref{fig:coverage}).
Considering only functions with type signatures compatible with Model (brown
bars in \Cref{fig:coverage}), we were able to synthesize implementations for
nearly 50\% on average. This represents a significant proportion of each
library's behavior---even in our worst performing case (BLAS), we are able to
synthesize nearly 20\% of all functions in the library. Performance on the BLAS
library is limited by the high complexity of many of its constituent functions
(e.g.\ solving systems of equations on packed matrix structures).

For each synthesis failure in our set of evaluation functions, we examined the
reference function by hand to determine why it could not be synthesized. In some
cases (e.g.\ \texttt{strtok} from \texttt{string.h}), the function demonstrated
stateful behavior. Modeling this type of function is an open problem in
program synthesis, with recent work addressing limited contexts such as heap
manipulation \cite{Polikarpova2019a}. Our synthesis methodology presumes that
target functions are idempotent, and so does not support stateful functions.
Doing so is interesting future work. A small number of functions (e.g.\
\texttt{ssyr2k} from \texttt{blas}) exhibit unusual control flow idioms not
expressible using our set of fragments. However, the majority of failures are
timeouts resulting from long required sequences of instructions in target
functions.

\pgfkeys{/pgf/fpu=true}
\pgfkeys{/pgf/number format/.cd,fixed,fixed zerofill,precision=1}

\begin{figure}[t]
  \centering
  \includegraphics[width=0.85\columnwidth]{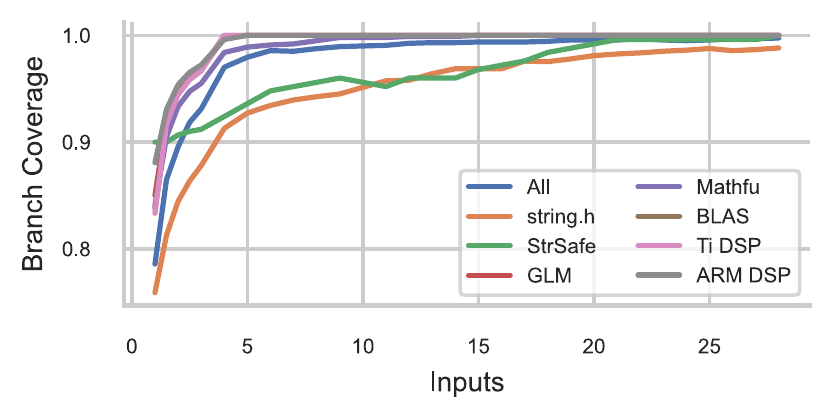}

  \caption{
    Corpus branch coverage achieved using randomly generated inputs. Coverage
    values are reported as the mean of three separate runs.
  }

  \label{fig:branchcoverage}
\end{figure}

Program synthesis over an entire library API is a challenging problem; the
programs that we were able to synthesize are considerably more complex than
comparable work in program synthesis while requiring less information to do so.
\cite{So2017,Rosin2018}.

\subsubsection{Difficulty}

\Cref{fig:difficulty} shows the distribution of candidate functions evaluated
for the synthesis problems in each library. From these distributions we see that
the majority (84\%) of functions were synthesized in less than 2 minutes.  We
were able to evaluate approximately 1,000 candidates per second on an 8-core
desktop-class machine.

The distribution of synthesis times is long-tailed; only two functions from the
\texttt{BLAS} library took more than 2 hours to synthesize. These synthesis
times are comparable to existing work in program synthesis, and could be further
improved by using techniques such as hill-climbing to guide the search process.

\subsubsection{Comparison to TypeDirect}

We were only able to identify one other program synthesizer with library
functions and migrations as an explicitly stated goal. In that work,
partial semantic knowledge and type information is used to guide a synthesizer
(\textsc{TypeDirect}) towards synthesized implementations of performance
bottleneck functions \cite{Collie2019}

Evaluation of \textsc{TypeDirect} is limited to 12 such functions, with
synthesis guided by annotations that specify semantic properties of the target
functions \cite{Collie2019}. We restated our set of synthesis problems for
\textsc{TypeDirect} and recorded how many it could synthesize. These results are
compared to those achieved by Model in \Cref{fig:coverage} (green bars); Model
performs significantly better across all the libraries evaluated, with
\textsc{TypeDirect} failing to synthesize any function in four of the seven
libraries. Additionally, \textsc{TypeDirect} took longer to synthesize the
functions for which it was successful (up to 4 hours in some cases).

In sum, compared to TypeDirect, we find Model is (1) automated and easy to use,
not relying on annotations to guide synthesis and (2) more widely applicable
with better synthesis coverage across different libraries. This is due primarily
to \textsc{TypeDirect}'s focus on synthesis for specific accelerator libraries
rather than general API migration.

\subsection{RQ2: Correctness of Synthesis} \label{sec:rq2}

For every synthesized library function, we automatically generated random and
boundary value inputs and checked if outputs matched those from the target
black-box function. 

\paragraph{Random Testing.}

We generated test inputs for every synthesized candidate by uniformly sampling
values in the range of the input data types, as described in
Section~\ref{sec:randomgen}.  We found all the synthesized library
implementations were behaviorally equivalent to the target functions with
respect to the random inputs generated for them. 

\paragraph{Manual Check.} 

As well as testing using random IO examples, we examined each synthesized
solution manually using our knowledge of their intended behavior. Only one
program was judged to be incorrect: the \texttt{memmove} function from
\texttt{string.h}. If the memory regions passed as arguments aliased (i.e.\ they
overlapped), the synthesized implementation would exhibit incorrect behavior.
Our testing methodology did not generate aliased memory. We generated a set of
aliased inputs manually and were able to correctly synthesize \texttt{memmove}.

\paragraph{Boundary Value Testing.}

We additionally tested each synthesized candidate using boundary and outside
range values for inputs. In every case, the synthesized candidate conformed to
the expected behavior on these inputs.

\paragraph{Adequacy of testing.}

We assessed the adequacy of the generated inputs in exercising behaviors of the
synthesized implementations by measuring the branch coverage achieved.
\Cref{fig:branchcoverage} shows the branch coverage achieved across the full set
of library functions evaluated. With as few as 10 distinct inputs, more than
98\% of the branch choices in our corpus of synthesized programs are evaluated.
Typically, at most around 30 random inputs are needed to provide 100\% branch
coverage for a synthesized candidate. The numerical libraries we evaluate most
often contain loops as their primary control flow; branch coverage is less
difficult to achieve over looping code than over conditionals.

These results provide confidence that the synthesized candidates behave
equivalently to the target program with respect to inputs that exercise the
complete control flow in the candidates.  

\paragraph{Inside the Black Box}

In many cases, we had the source code for libraries, making it possible to
directly compare our synthesized programs to the original code by ``looking
inside'' the black-box. These programs were compiled to LLVM IR and used as
input to the Match and Migrate phases as if they had in fact been synthesized.
We did not identify any meaningful divergence in results; we achieved similar
per-library branch coverage, and the compiled IR for synthesized and handwritten
implementations was almost identical in most cases. No behavioral differences
were observed.

\subsection{RQ3: Accuracy of Match} \label{sec:rq2results}

\begin{table}[t]
  \centering

  \caption{
    Number of call sites where synthesized functions were inlined in each
    application, along with the proportion of these that were successfully
    rediscovered using Match.
  }

  \label{tab:match}
  \begin{tabular}{l|rr|r}
  \toprule

  \multirow{2}{*}{\textbf{Application}} & \multicolumn{2}{c|}{\textbf{Inlined Calls (\LL)}} & \textbf{\# User Code} \\
                                        & \#Instances & Matched (\%) & \textbf{Matches} (\textbf{C$\to$L}) \\
  \midrule

  \texttt{ffmpeg}       & 4,976 & 100\% & 24 \\
  \texttt{texinfo}      & 586   & 100\% & 1 \\
  \texttt{xrdp}         & 686   & 100\% & 0 \\
  \texttt{coreutils}    & 623   & 100\% & 16 \\
  \texttt{gems}         & 46    & 100\% & 61 \\
  \texttt{darknet}      & 128   & 100\% & 13 \\
  \texttt{caffepresso}  & 189   & 100\% & 0 \\
  \texttt{nanvix}       & 0     & 100\% & 16 \\
  \texttt{etr}          & 4     & 100\% & 45 \\
  \texttt{androidfs}    & 0     & 100\% & 2 \\

  \midrule

  \textbf{Total} & \textbf{7,238} & & \textbf{178} \\
  \bottomrule
  \end{tabular}
\end{table}

We assessed if the constraint descriptions of every synthesized library
implementation was able to match the inlined implementation of the same library
in application code (\LL{}). Match is able to successfully identify every
instance of inlined code across all the applications we evaluated; the number of
inlined instances for each application is given in \Cref{tab:match}. This is
because the same code is inlined at each site, and because inlining does not
change the structure of the code from which the constraint description was
generated.

As well as being able to successfully identify inlined calls, Match is able to
identify locations in the application code where equivalent functionality to a
library function is implemented, \textbf{C$\to$L} (number of instances shown in
\Cref{tab:match}). We performed a manual search for further instances not
discovered by Match based on these results. A combination of several techniques
was used to perform this search: we used handwritten CAnDL constraints for
significantly abstracted versions of each function to guide an initial search,
as well as textual similarity and heuristic exploration of the code.

For example, where a re-implementation of one string-processing function was
discovered, we searched by hand for similar re-implementations that were not
discovered by Match. For a region to be classified as a re-implementation, we
required that on well-formed inputs (i.e.\ not accounting for ``exceptional''
control flow), the region performs the same task as the original function.

No further instances of this kind were identified by this search, confirming
with reasonable certainty that there were no false negatives from Match (though
no technique can verify this formally). The constraints generated by Match were
specific enough that none of the application code matches represented false
positives.

Running the CAnDL solver takes additional time during compilation;
approximately the same as compilation itself for each pattern to be searched
for \cite{Ginsbach2018a}. This time is not a bottleneck when using \mmm{}
practically.

\subsection{RQ4: Accuracy of Migrate} \label{sec:rq3results}

\begin{table}[t]
  \centering

  \caption{
    Migration opportunities discovered in each application, broken down by the
    category of the source context (source library calls \textbf{L} or user code
    \textbf{C}).
  }

  \label{tab:migrate}

  \begin{tabular}{l|r|rrr}
  \toprule

  \multirow{2}{*}{\textbf{Application}} & 
  \multirow{2}{*}{\textbf{Migrations}} & 
  \multicolumn{3}{c}{\textbf{Category}} \\
                                        & 
                                        &
  \LLPr{} & \CL{} & \LCL{} \\
  \midrule

  \texttt{ffmpeg}       & 655 & 629 & 24 & 2 \\
  \texttt{texinfo}      & 431 & 413 & 1 & 17 \\
  \texttt{xrdp}         & 274 & 269 & 0 & 5 \\
  \texttt{coreutils}    & 649 & 633 & 16 & 0 \\
  \texttt{gems}         & 107 & 46 & 61 & 0 \\
  \texttt{darknet}      & 40 & 7 & 13 & 20 \\
  \texttt{caffepresso}  & 24 & 24 & 0 & 0 \\
  \texttt{nanvix}       & 16 & 0 & 16 & 0 \\
  \texttt{etr}          & 49 & 4 & 45 & 0 \\
  \texttt{androidfs}    & 2 & 0 & 2 & 0 \\
  \midrule

  \textbf{Total} & \textbf{2,247} & \textbf{2,025} & \textbf{178} & \textbf{44} \\
  \bottomrule
  \end{tabular}
\end{table}

For every synthesized target library function, we assessed in how many cases the
generated constraints for that function matched application code that was not
originally a call to that function. This quantifies the number of possible
migrations enabled by \mmm{}. \Cref{tab:migrate} gives the total number of
migrations found in each application, as well as a breakdown into three
categories:
\begin{itemize}
  \item Replacement of a source function with a semantically identical target
    function from a different library (\LLPr{}).

  \item Identification and replacement of redundant application code that could be
    better expressed as a target library function call (\CL{}).

  \item Replacement of code that \emph{combines} a source library call and
    handwritten code with a target function (\LCL{}).
\end{itemize}

The most common migrations were \LLPr{}, where two libraries implemented the
same function (for example, delimited string copying or a vector dot product).
Some functions did not produce migration opportunities, even though they could
be inlined and matched. \texttt{memcpy} is an example of this; applications like
\texttt{ffmpeg} and \texttt{xrdp} that frequently perform buffer copies show far
fewer migrations than inlined matches.

Note that the category \CL{} corresponds exactly to the number of matches in
user code (\textbf{C$\to$L}) quoted in \Cref{tab:match}. This is because any
matching instance of a function in application code represents a migration
opportunity; there is no original function whose matches we are not interested
in.

These results demonstrate that \mmm{} is able to successfully identify distinct
classes of migration (other tools are often limited to one of these classes
only, and \LCL{} migrations generally require expert knowledge to express). The
migrations we identify are useful and would be difficult to identify with
existing tools.

\subsection{Threats to Validity}

We find \mmm{} is able to identify and perform a large number of useful
migrations using real-world applications and libraries, in contexts not well
served by existing tools. The primary threats to \emph{internal validity} are:
(1) The fragment vocabulary used by Model is a limiting factor; the variety of
programs that can be synthesized depends on this vocabulary. However, this is a
limitation shared by all sketching program synthesizers. (2) Our CAnDL
constraint generation is not formally verified; we rely on testing with
different library functions to check constraints always match their source
programs. (3) We check the correctness of the synthesized implementations
against target functions using test inputs that achieve branch coverage.
Proving total correctness is known to be challenging~\cite{deville1994logic},
especially when the target source code is not visible.

The main threat to \emph{external validity} lies in the subject libraries chosen
and the restriction to two problem domains: string processing and mathematical
operations. These domains have also been targeted by other migration tools and
we used these to facilitate comparison. Our synthesis technique is not
restricted to these domains and we will apply our techniques to other domains in
the future; extending the vocabulary of fragments to include more expressive
computations will allow us to scale synthesis to more complex APIs and
functions.

\section{Related Work} \label{sec:related}
\subsection{Library Migration}

(Semi-)automatic rewriting of application code to use new libraries has been
well studied, particularly for Java and other object-oriented languages
\cite{zaidi2019library,xavier2017historical}. \citet{Robillard2013} partition
migration techniques into 3 sub-areas: library upgrade \cite{Xing2007,Dig2006},
API evolution \cite{Schafer2008,Dagenais2008} and library migration
\cite{Wu2010,Zhong2010,Nguyen2010}. Many schemes rely on a large corpus of
programs using the old and new libraries, frequently focusing on change logs
\cite{Teyton2013a}. This ongoing need is highlighted by \citet{Alrubaye2018},
whose work aims to automatically identify key changes that produce a migration.

\subsubsection{Automatic}

Different levels of abstraction delineate automatic approaches. Similarity of
text description has been used to map old to new APIs
\cite{pandita2015discovering}, while others
\cite{Phan2017a,nguyen2016api,nguyen2017exploring} use a syntactic view of
programs to build a learned vector-space encoding \cite{Mikolov2013} for
migration given an initial parallel mapping. Although the embedding-based
approach taken by API2Vec \cite{nguyen2017exploring} is flexible, the resulting
ambiguity is in fact a hindrance when performing migrations. More recent work
attempts to generate mappings between APIs based on their usage
\cite{chen2019mining}.

Other work \cite{xu2019meditor, yang2018edsynth} goes beyond simple replacement
of library API calls. \citet{xu2019meditor} use syntactic program differencing
and program dependency analysis to target actual edits and replacements.
Although it is a syntactic rather than semantic approach, they are able to add
new code to help migration of libraries. \textsc{EdSynth} \cite{yang2018edsynth}
synthesizes candidate API calls to fill partial program using information from
test executions and method constraints. Unlike our synthesis approach, their
work requires white-box information on candidate methods, exact locations to
insert API calls, and a user-provided test suite to serve as a correctness
specification.

Closer to our aim of not relying on prior API mapping examples is the approach
taken by \citet{Bui2019}. It uses GANs to generate initial migrations (seeds)
rather than using human knowledge to do so \cite{bui2019sar}. To achieve this,
it makes the assumption that use of APIs when migrating remains roughly the
same. It has significantly lower precision than our approach, relies on lexical
similarity and cannot perform \CL{} migrations. Other work uses specific
semantic knowledge of functions to perform refactoring with semantic guarantees
\cite{Shaw2014}.

\subsubsection{Expert-Driven}

A different approach to API migration is to use expert knowledge to encode
migration patterns by hand, then compile them to a searchable representation to
perform migrations. This approach is taken by tools like ReFaster (Java)
\cite{Wasserman2013} and NoBrainer (C / C++) \cite{Savchenko2019}; they permit
complex migrations but require experts to create the migration patterns
initially. Similarly, IDL \cite{Ginsbach2018} implements migrations of
computational ``idioms'' to target heterogeneous computing platforms. The
underlying code search mechanism for \mmm{} (CAnDL \cite{Ginsbach2018a}) can be
used to implement this style of migration tool in a portable way. \mmm{} extends
prior work by adding a \emph{learning} phase that creates migration patterns
automatically.

\subsection{Program Synthesis}

\mmm{} uses program synthesis as a technique for modeling the behavior of
library functions. We give a brief overview of related work in synthesis.

Prior work in imperative synthesis frequently focuses on straight-line code
\cite{Gulwani2011a, Sasnauskas2017} or has to make special provision for
control-flow \cite{Gulwani2011}. \textsc{Simpl} overcomes this problem by
assuming a partial program is already provided (such as a loop structure)
\cite{So2017}. Other
work aims to complete suggested sketches \cite{solar2013program} of programs to
provide programmer abstraction and auto-parallelization
\cite{Fedyukovich:2017:GSS:3062341.3062382}

Type signatures and information are often used to direct program synthesis, most
commonly for functional programs \cite{Osera2015,Osera2019}. Other work uses
extended type information as a means of accessing heterogeneous accelerators
\cite{Collie2019}. Our work considers a much wider, more diverse class of
libraries and applications without additional annotation.

Others have used neural components to improve the performance of an existing
synthesizer. For example, both \textsc{DeepCoder} \cite{Balog2016} and
\textsc{PCCoder} \cite{Zohar2018a} aim to learn from input-output examples; both
require fixed-size inputs and outputs and use a small DSL to generate training
examples. Learned programs are limited to list processing tasks; the DSLs
targeted by these (and similar implementations such as \textsc{SketchAdapt}
\cite{Nye2019}) also rely on high level primitive including (for example)
primitives to tokenize strings or perform list manipulations. 

Operating under the assumption of a black-box API  means that many existing
approaches in program synthesis do not apply or fail to generalize to our
context \cite{Chen2019,Chen2019a}. By using a black-box oracle we are able to
avoid issues of generalization across datasets \cite{Parisotto2016a,Kalyan2018}.

\section{Conclusion} \label{sec:conclusion}
In this paper we have proposed a novel API migration problem that matches
real-world problem contexts. Our approach, \mmm{}, uses the behavior of library
functions to discover migrations without expert knowledge, changelogs, or
access to the library's source code.

We successfully applied this approach to 7 large, widely-used libraries and were
able to successfully synthesize nearly 40\% of their functions. We discovered
over 7,000 instances of these functions in 10 well-known C/C++ applications, and
were able to discover a number of missed optimizations and bugs.

Using constraint-based search for API migration allows for the \emph{semantics}
of the code in question to be accounted for, rather than just the contexts in
which they appear; this results in more precise migrations. Future work applying
these methods to more domains is likely to be interesting.

\begin{acks}
  This work was supported by the \grantsponsor{epsrc}{Engineering and Physical
  Sciences Research Council}{http://https://epsrc.ukri.org/} (grant
  \grantnum{ppar}{EP/L01503X/1}), EPSRC Centre for Doctoral Training in
  Pervasive Parallelism at the University of Edinburgh, School of Informatics.
\end{acks}

\clearpage

\bibliographystyle{ACM-Reference-Format}
\bibliography{bibliography}

\end{document}